# Constraints on the density dependence of the symmetry energy

M.B. Tsang(曾敏兒)[1,2,*], Yingxun Zhang(张英逊)[1,3], P. Danielewicz[1,2], M. Famiano[4], Zhuxia Li(李祝霞)[3], W.G. Lynch(連致標)[1,2], A. W. Steiner[1,2]

[1]*Joint Institute of Nuclear Astrophysics and National Superconducting Cyclotron Laboratory, Michigan State University, East Lansing, MI 48824, USA.*
[2]*Physics and Astronomy Department, Michigan State University, East Lansing, MI 48824, USA.*
[3]*China Institute of Atomic Energy, P.O. Box 275 (18), Beijing 102413, P.R. China.*
[4]*Physics Department, Western Michigan University, Kalamazoo, MI, USA.*

## Abstract

Collisions involving $^{112}$Sn and $^{124}$Sn nuclei have been simulated with the Improved Quantum Molecular Dynamics transport model. The results of the calculations reproduce isospin diffusion data from two different observables and the ratios of neutron and proton spectra. By comparing these data to calculations performed over a range of symmetry energies at saturation density and different representations of the density dependence of the symmetry energy, constraints on the density dependence of the symmetry energy at sub-normal density are obtained. Results from present work are compared to constraints put forward in other recent analyses.

---

- Corresponding author: tsang@nscl.msu.edu

The nuclear symmetry energy, which describes the difference between the binding energy of symmetric matter, with equal proton (Z) and neutron (N) numbers, and that of pure-neutron matter, governs important properties of nuclei and neutron stars [1,2]. In nuclei, the symmetry energy and the Coulomb energy define the binding energy minimum where nuclei are stable against beta decay [3,4]. In neutron stars, the symmetry energy and its density dependence influence strongly the nature and stability of the phases within the star [5], the feasibility of direct URCA cooling processes within its interior [6], the composition [5] and the thickness of its inner crust, the frequencies of its crustal vibrations, as well as its radius [7].

The symmetry energy contribution to the energy per nucleon in uniform matter can be written as $E_\delta = S(\rho)\delta^2$, where the asymmetry $\delta = (\rho_n - \rho_p)/\rho$ and $\rho_n$, $\rho_p$ and $\rho$ are the neutron, proton and nucleon number densities, and $S(\rho)$ describes the density dependence of $E_\delta$. Summing the contributions from the symmetry energy and the energy per nucleon of symmetric matter, $E(\rho, \delta) = E_0(\rho, \delta=0) + E_\delta$, provides the total energy per nucleon (i.e. the Equation of State (EoS)) of cold matter. Both symmetry and total binding energy terms in the nuclear semi-empirical binding energy formulae reflect averages of $E_\delta$ over the densities of nuclei [3]. The values of surface and volume symmetry energy terms obtained over fits of such formula to measured masses provide some sensitivity to the density dependence of $S(\rho)$ near saturation density [3,4]. Giant Dipole Resonances, low-lying electric dipole excitations, charge exchange spin-dipole excitations and the difference between neutron and proton matter radii may also provide sensitivity to the density dependence of $S(\rho)$ down to around 0.1 $fm^{-3}$ [8-13].

Nuclear collisions can produce larger transient density variations. Significant constraints on the symmetric matter Equation of State at $1 \leq \rho/\rho_0 \leq 4.5$ have been obtained from measurements of collective flow [14] and Kaon production [15]. On the other hand, the symmetry energy has been recently probed at sub-saturation densities ($0.4 \leq \rho/\rho_0 \leq 1.2$) in collisions via measurements of isospin diffusion [16,17], and of double ratios involving neutron and proton energy spectra [18]. These two observables largely reflect the transport of nucleons under the influence of nuclear mean fields and of the collisions induced by the residual interactions, both of which can be described by transport theory. Using a Boltzmann-Uehling-Uhlenbeck (BUU) transport model IBUU04, Chen et al., obtained a reasonable description of isospin diffusion data with a symmetry energies of approximate form $S(\rho) \approx 31.6(\rho/\rho_0)^\gamma$ with γ=0.69-1.1 [19]. The double ratios of neutron and proton energy spectra [18], however, have not been adequately described by the IBUU04 model [20].

In this paper, we show that both the isospin diffusion and neutron to proton ratios observables can be described by the Improved Quantum Molecular Dynamics (ImQMD) transport model. Detailed description of the model and its application to the double ratio data can be found in ref. [21]. For brevity, we limit our discussion here to the parameterization of the symmetry energy used in our calculations, which is of the form

$$S(\rho) = \frac{C_{s,k}}{2}\left(\frac{\rho}{\rho_0}\right)^{2/3} + \frac{C_{s,p}}{2}\left(\frac{\rho}{\rho_0}\right)^{\gamma_i}, \qquad (1)$$

Unless indicated otherwise, the kinetic and potential parameters are $C_{s,k}$=25 MeV, $C_{s,p}$=35.2 MeV and the symmetry energy at saturation density, $S_0 = S(\rho_0)$ =30.1 MeV.

Both the QMD and BUU models calculate the trajectories of nucleons, taking into account the influence of self-consistent mean fields and nucleon-nucleon collisions via the residual interaction. The two models have been shown to provide similar predictions for one body observables [22]. However, fluctuations are averaged out by the parallel calculations in BUU involving test particles. In the QMD approach, the N-body equations for nucleons are solved event by event. This enhances the importance of fluctuations and correlations in QMD and provides a mechanism to calculate the production of complex nuclei. At incident energies of E/A=50 MeV where the present studies are conducted, cross sections for production of complex nuclei are significant and the influence of cluster production cannot be neglected [21]. This makes direct comparisons of data to QMD models more straightforward than for BUU models.

We first turn our attention to the interpretation of neutron/proton double ratio data within the ImQMD model [21]. This observable derives its sensitivity to the symmetry energy from the opposite sign of the symmetry potential for neutrons as compared to protons [23]. First experimental comparisons of neutron to proton spectra in ref. [18] used a double ratio in order to reduce sensitivity to uncertainties in the neutron detection efficiencies and sensitivity to relative uncertainties in energy calibrations of neutrons and protons. This double ratio,

$$DR(Y(n)/Y(p)) = R_{n/p}(A)/ R_{n/p}(B) = \frac{dM_n(A)/dE_{C.M.}}{dM_p(A)/dE_{C.M.}} \cdot \frac{dM_p(B)/dE_{C.M.}}{dM_n(B)/dE_{C.M.}}, \quad (2)$$

is constructed from the ratios of energy spectra, $dM/dE_{C.M.}$ of neutrons and protons for two systems A and B characterized by different isospin asymmetries. The star symbols in the left panel of Figure 1 show the neutron-proton double ratios measured at $70° \leq \theta_{C.M.} \leq 110°$ as a function of center-of-mass (C.M.) energy of nucleons emitted from the central collisions of $^{124}$Sn+$^{124}$Sn and $^{112}$Sn+$^{112}$Sn [18].

We have performed calculations for two systems: A=$^{124}$Sn+$^{124}$Sn and B=$^{112}$Sn+$^{112}$Sn. About 60,000 events have been simulated for each impact parameter. Within the statistical uncertainties, the double ratio observable, DR(Y(n)/Y(p)), is nearly independent of impact parameter over a range of 1 fm $\leq$ b $\leq$ 5 fm. The lines in the left panel of Figure 1 show double ratios vs. the C.M. energy of nucleons for $\gamma_i$=0.35, 0.5, 0.75, 1 and 2 averaged over b=1, 2, 3 fm. The computation uncertainties in Figure 1 are statistical. Despite the large experimental uncertainties for higher energy data, these comparisons definitely rule out both very soft ($\gamma_i$=0.35, dotted line with closed diamond points) and very stiff ($\gamma_i$=2, dotted line with open diamond symbols) density-dependent symmetry terms. The right panel shows the dependence on $\gamma_i$ of the *total* $\chi^2$ computed from the difference between predicted and measured double ratios. We determine, to within a 2$\sigma$ uncertainty, parameter values of $0.4 \leq \gamma_i \leq 1.05$ corresponding to an increase in $\chi^2$ by 4 above its minimum near $\gamma_i$~0.7.

At sub-normal density, higher symmetry energy, which enhances the emission of neutrons, are associated with decreasing $\gamma_i$. Similarly DR(Y(n)/Y(p)) increase with decreasing $\gamma_i$. However, in the limit of very small $\gamma_i \ll 0.35$ that the system completely disintegrates, DR(Y(n)/Y(p)) decreases and approaches the limit of $N_{total}/Z_{total}$ =1.2. As a consequence of these two competing effects, the double ratio values peak around $\gamma_i$~0.7.

The density dependence of the symmetry energy has also been probed in peripheral collisions between two nuclei with different isospin asymmetries by examining the diffusion of neutrons and protons across the neck that joins them. This "isospin diffusion" generally continues until the two nuclei separate or until the chemical potentials for neutrons and protons in both nuclei become equal. To isolate diffusion effects from other effects, such as pre-equilibrium emission, Coulomb effects and secondary decays, measurements of isospin diffusion compare "mixed" collisions, involving a neutron-rich nucleus A and a neutron-deficient nucleus B, to the "symmetric" collisions involving A+A and B+B. The degree of isospin equilibration in such collisions can be quantified by rescaling the isospin observable X according to the isospin transport ratio $R_i(X)$ [16]

$$R_i(X) = 2\frac{X - (X_{A+A} + X_{B+B})/2}{X_{A+A} - X_{B+B}}. \qquad (3)$$

where X is the isospin observable. In the absence of isospin diffusion, the ratios are $R_i(X_{A+B}) = R_i(X_{A+A}) = 1$ and $R_i(X_{B+B}) = R_i(X_{B+A}) = -1$. If isospin equilibrium is achieved, then the ratios $R_i(X_{A+B}) = R_i(X_{B+A}) \approx 0$ for the mixed systems.

Eq. (3) dictates that different observables, X, provide the same results if they are linearly related [17]. Experimental isospin transport ratios obtained from isoscaling parameters, α, [16] and from yield ratios of A=7 mirror nuclei, $R_7 = R_i(X_7=ln(Y(^7Li)/Y(^7Be)))$ are consistent, i.e. $R_i(\alpha) \cong R_7$, reflecting linear relationships between α, $X_7$, and the asymmetry δ of the emitting source [17]. This relationship also holds for fragments produced after sequential decays [24]. For emission at a specific rapidity y, we assume $R(\alpha)=R(\delta)=R_7$ to be valid, as has been confirmed experimentally [17,24] and theoretically for all statistical and dynamical calculations [25-27]. BUU calculations use the asymmetry of the projectile residues [16] to construct $R_i(\delta)$. In the following, we calculate δ from the asymmetry of the fragments and free nucleons emitted at the relevant rapidity, but our conclusions do not significantly change if fragments alone are used to calculate δ.

Experimental isospin diffusion transport ratios, $R_i(\alpha)$, plotted as shaded regions in the left panel of Figure 2, have been obtained using Eq. (3) with isoscaling data from $^{124}$Sn+$^{112}$Sn (top region) and $^{112}$Sn+$^{124}$Sn (bottom region) scaled by the symmetric collisions of $^{124}$Sn+$^{124}$Sn, $^{112}$Sn+$^{112}$Sn [16]. We have performed ImQMD calculations at impact parameters of b=5, 6, 7, and 8 fm. The lines in the left panel of Figure 2 show the predicted isospin transport ratio $R_i(\delta)$ as a function of impact parameter b for $\gamma_i$=0.35, 0.5, 0.75, 1 and 2. At sub-saturation densities, larger symmetry energies (and faster equilibration) occur for smaller $\gamma_i$ values in isospin diffusion. Thus we see a monotonic decrease of $R_i(\delta)$ with decreasing $\gamma_i$. Cross-section estimates suggest that the isospin diffusion data cover the impact parameter range of 5.8 to 7.5 fm [17] and the data primarily reflects contributions from impact parameters of b=6 to 7 fm. We have performed the $\chi^2$ analysis for both impact parameters in the right panel of Figure 2. Compared to Fig. 1, the $\chi^2$ minima are lower; using the same $2\sigma$ criterion, the analysis brackets the region $0.35 \leq \gamma_i \leq 1.0$.

Unlike BUU type calculations, ImQMD calculations provide predictions for fragment yields as a function of rapidity. The star symbols in Fig. 3 represent measured values of $R_7$ obtained from the yield ratios of $^7$Li and $^7$Be [17] at b=6 fm, as shown by the lines in the left panel. This first calculation of the shapes and magnitude of the rapidity dependence of the isospin transport ratios $R_7$ reproduces the trends accurately. The corresponding $\chi^2$ analysis at b=6 (solid curve) and 7 (dashed curve) fm in the right panel displays sharp minima. Using the $2\sigma$ criterion adopted previously, the analysis favors the region $0.45 \leq \gamma_i \leq 0.95$.

Constraints on the exponent $\gamma_i$ depends on the symmetry energy at saturation density, $S_0 = S(\rho_0)$. Increasing $S_0$ has the same effect on the isospin transport ratio as decreasing $\gamma_i$. To compare our results to constraints obtained from nuclear masses and nuclear structure, we expand $S(\rho)$ around the saturation density, $\rho_o$,

$$S(\rho) = S_0 + \frac{L}{3}\left(\frac{\rho - \rho_o}{\rho_o}\right) + \frac{K_{sym}}{18}\left(\frac{\rho - \rho_o}{\rho_o}\right)^2 + \ldots \qquad (4)$$

where L and $K_{sym}$ are the slope and curvature parameters at $\rho_o$. For realistic parameterization of $S(\rho)$, $K_{sym}$ is strongly correlated to L [4]. As the second term in Eq. (4) is much larger than the third term, we believe L can be determined more reliably than $K_{sym}$. Furthermore, the slope parameter, $L = 3\rho_0 |dS(\rho)/d\rho|_{\rho_0} = [3/\rho_0]p_0$, is related to $p_0$, the pressure from the symmetry energy for pure neutron matter at saturation density. The symmetry pressure, $p_0$, provides the dominant baryonic contribution to the pressure in neutron stars at saturation density [11-13, 17].

We have preformed a series of ImQMD calculations at b=6 fm with different values of $\gamma_i$ and $S_0$ to locate the approximate boundaries in the $S_0$ and L plane that satisfy the $2\sigma$ criterion in the $\chi^2$ analysis of the isospin diffusion data. The two diagonal lines in Figure 4 represents estimates in such effort. Examination of the symmetry energy functional forms along these boundaries, where the diffusion rates are similar but $S_0$ and L are different, reveal that diffusion rates predominantly reflect the symmetry energy at $\rho \approx 0.5\rho_0$ and somewhat below. Even though the sensitivity of constraints on $S_0$ and L to differences between $m_n^*$ and $m_p^*$, and to the in-medium cross-sections have not been fully explored, BUU calculations for isospin diffusion show much more sensitivity to $S_0$ and L than to these other quantities [28].

The dashed, dot-dashed and solid lines centered around $S_0$=30.1 MeV in Fig. 4 represents L values consistent with the analysis presented in Figs. 1, 2, and 3, respectively. According to the Typel-Brown correlation, the range of L values obtained in our analysis corresponds to a variation in the skin thickness of $^{208}$Pb of about ±0.04 fm [29,2]. The vertical line at $S_0$ = 31.6 MeV depicts the range of L-values obtained in ref. [1,20] from comparisons of IBUU04 calculations to the measured isospin diffusion data in Fig. 2. Constraints from the isoscaling analyses with statistical models in ref. [30] are not included as the current analysis is problematic as discussed in ref. [31].

We have included, in Figure 4, other recent constraints in the density dependence of the symmetry energy. The lower box centered at $S_0$ = 32 MeV depicts the range of $p_0$ values from analyses of Pygmy Dipole Resonance (PDR) data [9]. The values of $p_0$ are given in the right axis. The upper box centered at $S_0$ = 32.5 MeV depicts the constraints reported in Ref. [4] from the analyses of nuclear surface symmetry energies. The shaded region in the inset of Figure 4 shows the density dependence of the symmetry energy of the shaded region bounded by $S_0$ = 30.2 and 33.8 MeV, the limiting $S_0$ values given by the PDR data [9]. The range $S_0$ adopted here is consistent with the finding from the charge exchange spin-dipole resonance result [10]. The Giant Dipole Resonances (GDR) result of ref. [8] is plotted as a solid circle in the inset.

In summary, both isospin diffusion and double ratio data involving neutron and proton spectra have been consistently described by a QMD model. The analyses of all three observables provide consistent constraints on the density dependence of the symmetry energy. The results overlap with recent constraints obtained from Giant Dipole Resonances, Pygmy Dipole Resonance and mass data. Some shifts in the boundaries of the constraints can be expected with improvements in the precision of the experimental data and in the understanding of theory. Nevertheless, the consistency between the different probes of the symmetry energy suggests that increasingly stringent constraints on the symmetry energy at sub-saturation density can be expected.

**Acknowledgement**

This work has been supported by the U.S. National Science Foundation under Grants PHY-0216783, 0456903, 0606007, 0800026, the High Performance Computing Center (HPCC) at Michigan State University, the Chinese National Science Foundation of China under Grants 10675172, 10175093, 10235030, and the Major State Basic Research development program under contract No. 2007CB209900.

**References**

[1] Bao-An Li, Lie-Wen Chen, Che Ming Ko, Phys. Rep. 464, 113 (2008).
[2] A.W. Steiner, M. Prakash, J.M. Lattimer, P.J. Ellis, Phys. Rep. 411, 325 (2005).
[3] P. Danielewicz, Nucl. Phys. A727, 233 (2003) and references therein
[4] Pawel Danielewicz and Jenny Lee, arXiv:0807.3743 and references therein
[5] A.W. Steiner, Phys. Rev. C 77, 035805 (2008)
[6] B.G. Todd-Rutel, J. Piekarewicz, Phys. Rev. Lett. 95, 122501 (2005).
[7] J.M. Lattimer, M. Prakash, Science 442, 109 (2007)
[8] Luca Trippa, Gianluca Colò, and Enrico Vigezzi, Phys. Rev. C 77, 061304(R) (2008)
[9] A. Klimkiewicz et. al., Phys. Rev. C 76, 051603(R) (2007)
[10] H. Sagawa et. Al., Phys. Rev. C 76, 024301 (2007)
[11] C. J. Horowitz and J. Piekarewicz, Phys. Rev. Lett. 86, 5647 (2001).
[12] J. Piekarewicz, Phys. Rev. C 69, 041301 (2004).
[13] Satoshi Yoshida and Hiroyuki Sagawa Phys. Rev. C 73, 044320 (2006).
[14] P. Danielewicz, R. Lacey, W.G. Lynch, Science 298, 1592 (2002).
[15] C. Fuchs, Prog. Part. Nucl. Phys. 56, 1 (2006).
[16] M.B. Tsang, et al., Phys. Rev. Lett. 92, 062701 (2004).
[17] T.X. Liu, et al., Phys. Rev. C 76, 034603 (2007).
[18] M.A. Famiano, et al., Phys. Rev. Lett. 97, 052701 (2006).
[19] B.A. Li, L.W. Chen, Phys. Rev. C 72, 064611 (2005).
[20] B.A. Li, L.W. Chen, G.C. Yong, W. Zuo, Phys. Lett. B634, 378 (2006).
[21] Y. Zhang, P. Danielewicz, M. Famiano, Z. Li, W.G. Lynch, M.B. Tsang, Phys. Lett. B664, 145 (2008) and references therein.
[22] J. Aichelin et al, Phys. Lett. B 224, 34 (1989)
[23] B.A. Li, C.M. Ko, Z.Z. Ren, Phys. Rev. Lett. 78, 1644 (1997).
[24] W.G. Lynch et al., private communication.
[25] M.B. Tsang, et al., Phys. Rev. Lett. 86, 5023 (2001).
[26] A.S. Botvina, O.V. Lozhkin, W. Trautmann, Phys. Rev. C 65, 044610 (2002).

[27] A. Ono, P. Danielewicz, W.A. Friedman, W.G. Lynch, M.B. Tsang, Phys. Rev. C 68, 051601(R), (2003).
[28] D. Coupland et al., private communication.
[29] S. Typel, B.A. Brown, Phys. Rev. C 64, 027302 (2001).
[30] D.V. Shetty, S.J. Yennello, G.A. Souliotis, Phys. Rev. C 76, 024606 (2007).
[31] S. R. Souza, et al, Phys. Rev. C 78, 014605 (2008).

Figure 1 (Color online): Left panel: Comparison of experimental double neutron-proton ratios [16] (star symbols), as a function of nucleon center of mass energy, to ImQMD calculations (lines) with different density dependencies of the symmetry energy parameterized by $\gamma_i$ in Eq. (1). Right panel: A plot of $\chi^2$ as a function of $\gamma_i$.

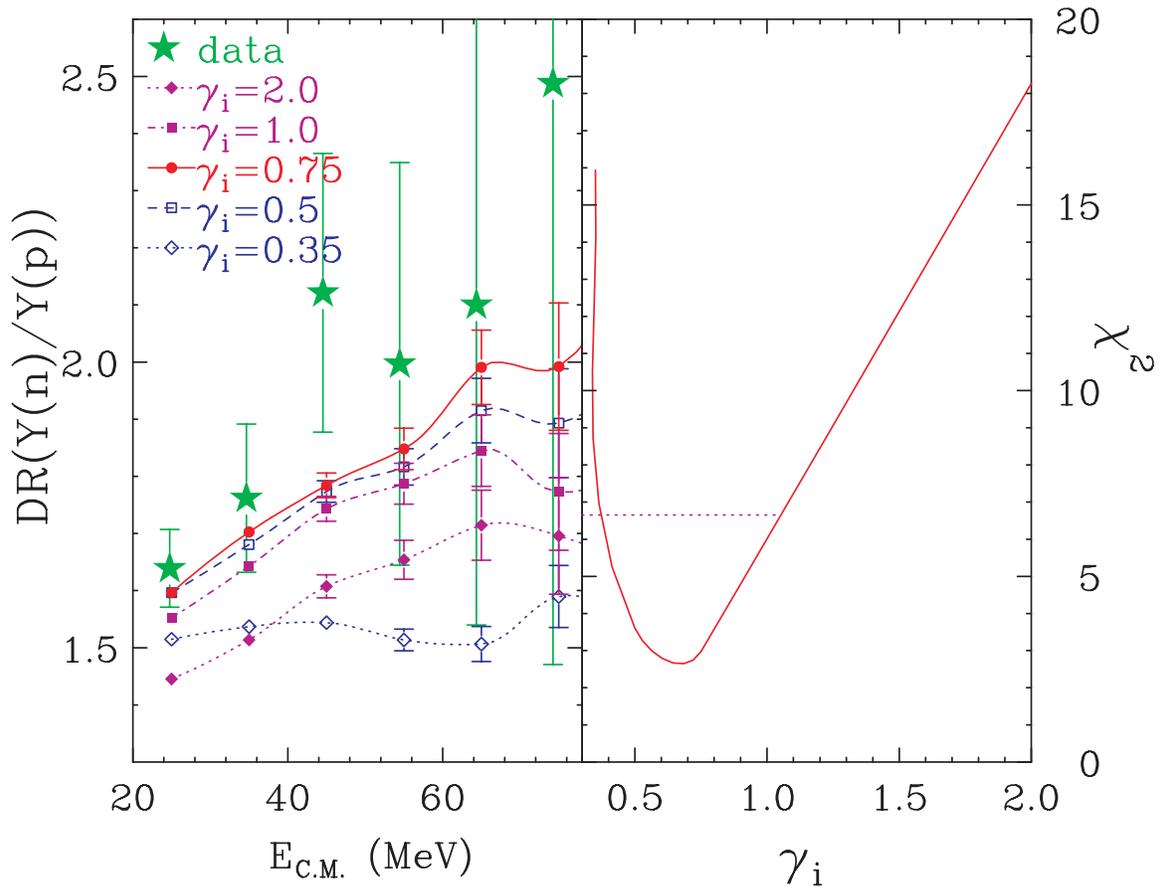

Figure 2 (Color online): Left panel: Comparison of experimental isospin transport ratios [14] (shaded regions) to ImQMD results (lines), for different impact parameters. Right panel: $\chi^2$ analysis for b=6 fm (solid curve) and b=7 fm (dashed curve) as a function of $\gamma_i$.

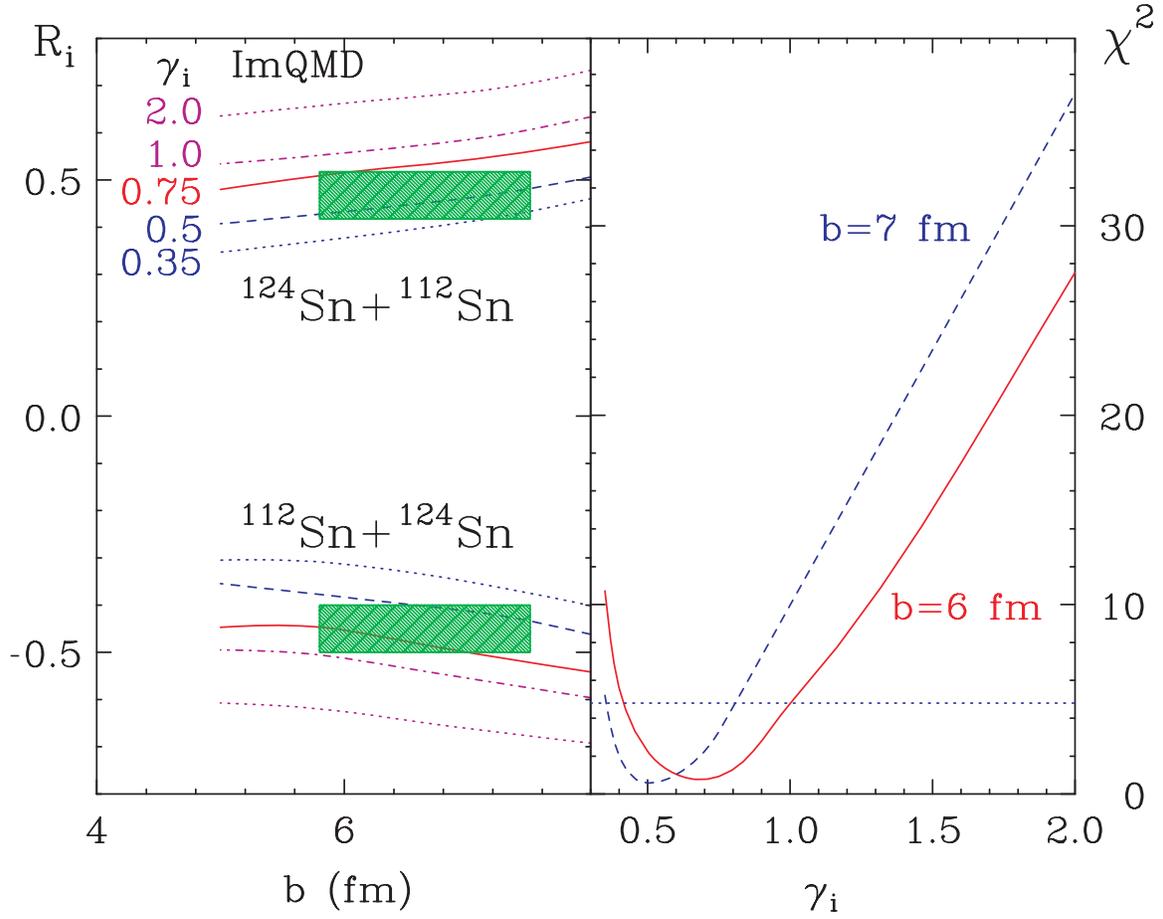

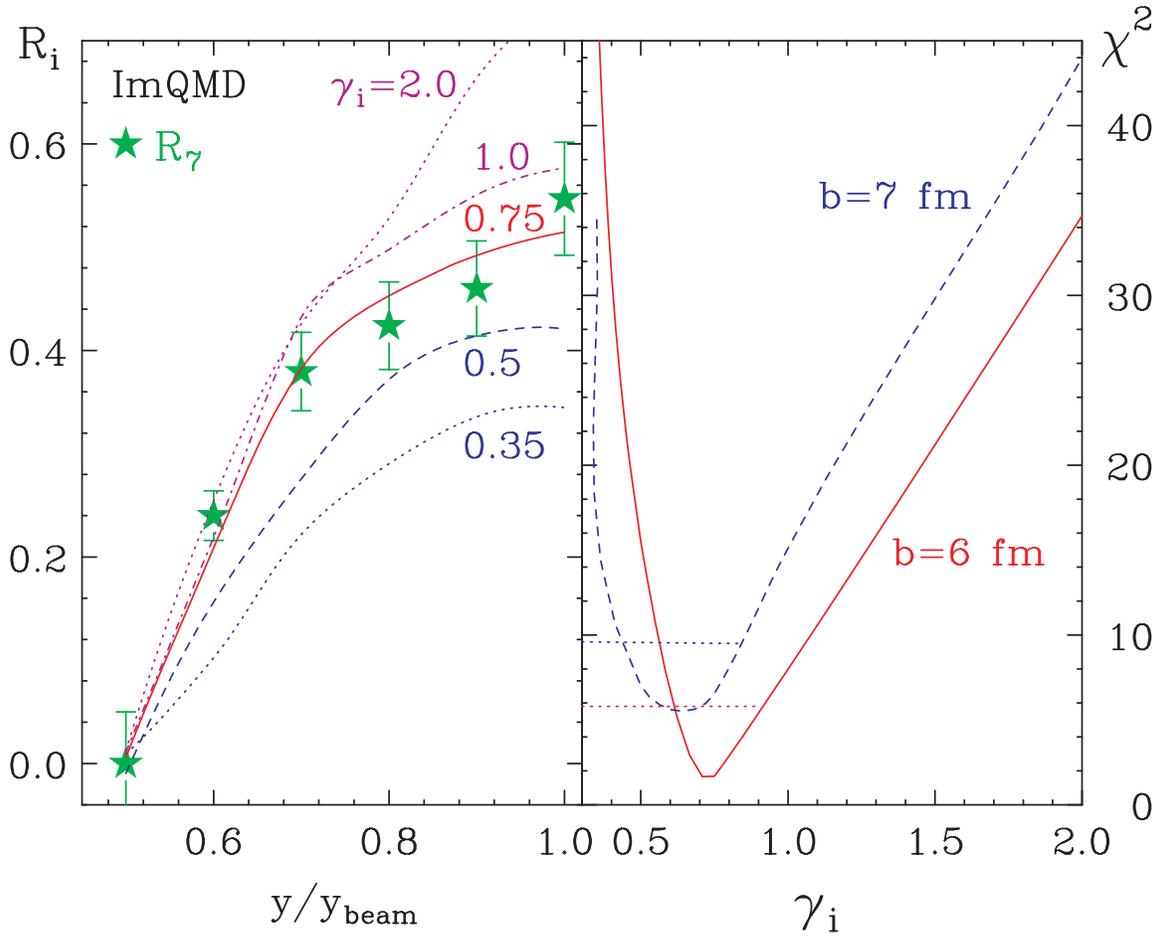

Figure 3 (Color online): Left panel: Comparison of experimental isospin transport ratios obtained from the yield ratios of A=7 isotopes [15] (star symbols), as a function of the rapidity, ImQMD calculations (lines) for b=6 fm. Right panel: $\chi^2$ analysis for b=6 fm (solid curve) and b=7 fm (dashed curve) as a function of $\gamma_i$.

Figure 4 (Color online): Representation of the constraints on parameters $S_0$ and L. The right axis corresponds to the neutron matter symmetry pressure at saturation density. The region bounded by the diagonal lines represents the constraints obtained in the present work. The vertical lines at $S_0 \sim 30.1$ MeV are obtained from the constraints stemming from Fig 1-3. The vertical line at $S_0 = 31.6$ MeV is from ref [1,18]. The lower and upper boxes are formed by the constraints from PDR data [9] and from symmetry energy analysis on nuclei [4], respectively. The inset shows the density dependence of the symmetry energy of the shaded region. The symbol in the inset represents the GDR results from ref. [8].

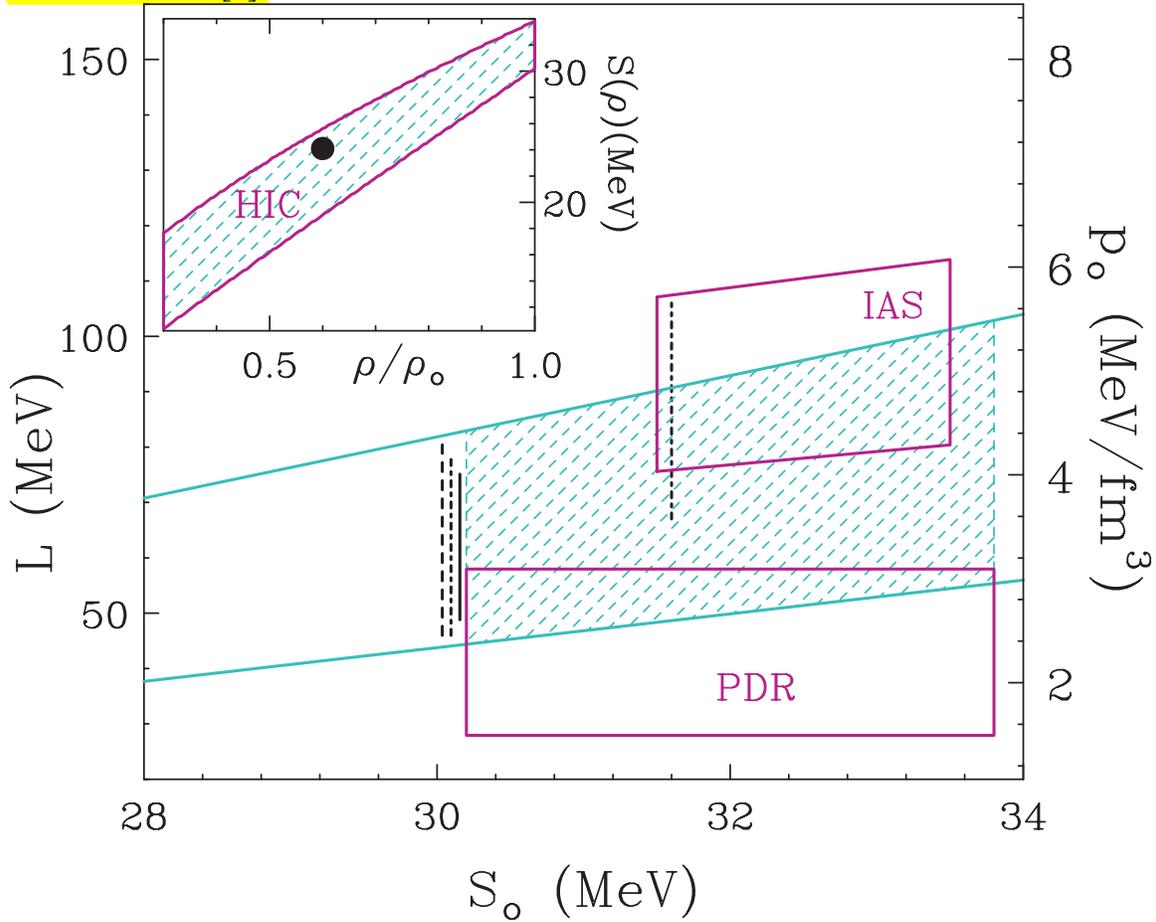